\documentclass[runningheads]{svmult}
\usepackage{makeidx}   % allows index generation
\usepackage{graphicx}  % standard LaTeX graphics tool
\usepackage{subeqnar}  % subnumbers individual equations
\usepackage{multicol}  % used for the two-column index
\usepackage{cropmark} % cropmarks for pages without
\usepackage{physprbb}  % centered layout of diverse elements, etc.
\def\lesssim{\mathrel{\hbox{\rlap{\hbox{\lower4pt\hbox{$\sim$}}}\hbox{$<$}}}}
\def\gtrsim{\mathrel{\hbox{\rlap{\hbox{\lower4pt\hbox{$\sim$}}}\hbox{$>$}}}}

\begin{document}

%\title*{ Ultra High Energy Neutrino-Neutrino Showers
% above GZK}
\title*{\center{ Shadows of Relic Neutrino Masses and Spectra  on Highest Energy GZK Cosmic Rays}}

\toctitle{UHE Neutrino- Relic Neutrino Imprints at GZK}

% allows explicit linebreak for the table of content
%
%
\titlerunning{UHE neutrino showering}
% allows abbreviation of title, if the full title is too long
% to fit in the running head
%
%%\author{Daniele Fargion\inst{1}
\author{D.Fargion,\inst {1}\ M.Grossi, \  P.G.De Sanctis Lucentini, \ C.Di
Troia,\ R.V.Konoplich \inst {2}}

\authorrunning{Daniele Fargion}
% if there are more than two authors,
% please abbreviate author list for running head
%
%
\institute { 1 Physics Department and INFN , Rome University 1,\\
     Pl.A.Moro 2, 00185, Rome, Italy\\
     2 Physics Dept.New York University,N.Y.,USA}

\maketitle              % typesets the title of the contribution

%\twocolumn

%\pagestyle{plain}

\begin{abstract}
The Ultra High Energy (UHE) neutrino scattering onto relic cosmic
neutrinos in galactic and local halos offers an unique way to
overcome GZK cut-off. The $UHE\,\nu$ secondary of $UHE$
photo-pion decays may escape the $GZK$ cut-off and travel on
cosmic distances hitting local light relic neutrinos clustered in
dark halos. The Z resonant production and the competitive
$W^+W^-$, ZZ pair production define a characteristic imprint on
hadronic consequent UHECR spectra. This imprint keeps memory both
of the primary $UHE\,\nu$ spectra as well as of the possible relic
neutrino masses values, energy  spectra and relic densities. Such
an hadronic showering imprint should reflect into spectra
morphology of cosmic rays near and above GZK ($10^{19} \, - \,
10^{21}$ eV) cut-off energies. A possible neutrino degenerate
masses  at $eVs$ or a more complex and significant neutrino mass
split  below or near Super-Kamiokande $\triangle m_{\nu_{SK}}\sim
0.1 eV $ masses might be reflected after each corresponding Z
peak showering, into new twin unexpected $UHECR$ flux modulation
behind $GZK$ energies: $ E_{p} \sim 3\left( \frac{\triangle
m_{\nu_{SK}}}{m_{\nu}} \right) \cdot 10^{21} \,eV $. Other extreme
shadows of lightest, nearly massless, neutrinos $
m_{\nu_{2K}}\simeq 0.001 eV \simeq \ kT_{\nu}$, their lowest
relic temperatures, energies and densities might be also
reflected at even higher energies edges near Grand Unification: $
E_{p} \sim 2.2\left( \frac{\ m_{\nu_{2K}}}{E_{\nu}} \right) \cdot
10^{23} \,eV $.
\end{abstract}

\section{Introduction}

Modern astro-particle physics face the old standing problem of
dark matter nature in galaxies up to cosmic scales. Neutrino with
a light mass may play a relevant role in solving the puzzle within
a hot-cold dark matter (HCDM) scenario. Moreover, at the edge of
highest energy astrophysics, the main open question regards the
nature of highest (Ultra High Energy, UHE) cosmic rays above the
Greisen Zatsepin Kuzmin cut-off ($\gtrsim 4 \cdot 10^{19}\,eV$).\\
These rare events almost  in isotropic spread are probably
originated by blazars AGN, QSRs  or GRBs in standard scenario,
and they should not come, if originally of hadronic nature, from
large distances because of the electromagnetic "dragging
friction" of cosmic 2.75 K BBR and of the lower energy diffused
inter-galactic radio backgrounds. Indeed as noted by Greisen,
Zatsepin and Kuzmin \cite{G},  \cite{ZK}, proton and nucleon mean
free path at E $> 5 \cdot 10^{19} \,EeV$ is less than 30 $Mpc$ and
asymptotically nearly ten $Mpc$.; also gamma rays at those
energies have even shorter interaction length ($10 \,Mpc$) due to
severe opacity by electron pair production via microwave and radio
background interactions \cite{Proth1997} . Nevertheless these
powerful sources (AGN, Quasars, GRBs) suspected to be the unique
source able to eject such UHECRs, are rare  at nearby distances
(($\lesssim 10 \div 20 \, Mpc$) as for nearby $M87$ in Virgo
cluster); moreover there  are not nearby $AGN$ in the observed
UHECR arrival directions.  Strong and coherent galactic (Biermann
1999-2000) or extragalactic (Farrar and Tvi Piran 1999-2000)
magnetic fields, able to bend such UHECR (proton, nuclei)
directions are not really at hand. The needed coherent lengths
and strength are not easily compatible with known cosmic data on
polarized Faraday rotation. Finally in latter scenario the same
contemporaneous  ultra-high energy $ZeV$ neutrons born, by
photo-pion production on BBR, may escape the magnetic fields
bending  and should keep memory of the primordial nearby ( let
say $M87$) arrival direction, leading to (unobserved)
in-homogeneities toward the primary source. Finally secondaries
EeV photons (by neutral pion decays) should also abundantly point
and cluster toward the same nearby $AGN$
sources \cite{El},\cite{Sigl}  contrary to (never observed) $AGASA$ data.\\
Another solution of the present GZK puzzle, the Topological
defects ($TD$), assumes as a source, relic heavy particles of
early Universe; they are imagined diffused as a Cold Dark Matter
component, in galactic or Local Group Halos . Nevertheless the
$TD$ fine tuned masses and ad-hoc decays are unable to explain the
growing evidences of doublets and triplets clustering in $AGASA$
$UHECR$ arrival data. In this scenario there have been recent
suggestions and speculations \cite{Blasi} for an unexpected
population of such 500 compact dark clouds of $10^8 M_{\odot}$,
each one made by such dark $TD$ clusters, spread in our galactic
halo ; they are assumed, nevertheless,  not correlated to luminous
known galactic halo, disk, globular clusters and center
components. We found all these speculations unnatural and not
plausible. On the other side there are possible evidences   of
correlation between UHECR arrival directions with far Compact
Radio Loud Quasar at cosmic distance (above GZK cut-off) ( Amitabh 2000).\\
Therefore the solution of UHECR puzzle based on primary Extreme
High Energy (EHE) neutrino beams(from AGN) at $E_{\nu} > 10^{21}$
eV and their undisturbed propagation from cosmic distances up to
nearby calorimeter made of  relic light $\nu$ in dark galactic or
local dark halo (Fargion, Salis 1997;Fargion,Mele,Salis 1999,
Weiler 1999, Yoshida et all 1998) is still, in our opinion, the
most favorite conservative solution for the GZK puzzle.
Interestingly new complex scenarios are then opening.\\

\section{UHE neutrino scattering in the halo: the three  neutrino masses, interaction scenarios}

\ If relic neutrinos have a mass around  an eVs they may  cluster
in galactic or Local Group halos, their scattering  with incoming
EHE neutrinos determine high energy particle cascades which could
contribute or dominate the observed UHECR flux at $GZK$ edges.
Indeed the possibility that neutrino share a little mass has been
reinforced by Super-Kamiokande evidence for atmospheric neutrino
anomaly via $\nu_{\mu} \leftrightarrow \nu_{\tau}$ oscillation.
Consequently there are at least two main extreme scenario for hot
dark halos: either $\nu_{\mu}\, , \, \nu_{\tau}$ are both
extremely light  ($m_{\nu_{\mu}} \sim m_{\nu_{\tau}} \sim
\sqrt{(\Delta m)^2} \sim 0.07 \, eV$) and therefore hot dark
neutrino halo is very wide, possibly degenereted (Gelmini 2000)
and spread out to local group clustering sizes (increasing the
radius but loosing in the neutrino density clustering contrast),
or $\nu_{\mu}, \nu_{\tau}$ have degenerated ($eV$ masses) split
by a very tiny different value. \\ In the latter fine-tuned
neutrino mass case ($m_{\nu}\sim 0.4 eV-1.2 eV$) (see Fig,2 and
Fig.3) the Z peak $\nu \bar{\nu}_r$ interaction (Fargion, Salis
1997;Fargion,Mele,Salis 1999, Weiler 1999, Yoshida et all 1998)
will be the favorite one while in the second case for heavier non
constrained neutrino mass ($m_{\nu} \gtrsim 5 \, eV$) only a $\nu
\bar{\nu}_r \rightarrow W^+W^-$ (Fargion,Mele,Salis 1999), and
the additional $\nu \bar{\nu}_r \rightarrow ZZ$ interactions,(see
the cross-section in Fig.1) considered here  for the first time ,
will be the only ones able to solve the GZK puzzle. Indeed the
relic neutrino mass within HDM models in galactic halo near
$m_{\nu}\sim 4 eV$ , corresponds to a "lower" and $Z$ resonant
incoming energy

\begin{subeqnarray}
%\sqrt{(\Delta m)^2}
 {{E_{\nu} =  {\left(
\frac{4eV} {\sqrt{{{m_{\nu}}^2+{p_{\nu}^2}}}} \right)} \cdot
10^{21} \,eV.} \nonumber}
\end{subeqnarray}

   This resonant incoming neutrino energy is able to shower only a
   small energy fraction into nucleons ($p,\bar{p}, n, \bar{n}$),
   (see $Tab.1$ below), at energies $E_{p}$ quite below GZK cut-off (see $Tab.2$
   below).

\begin{subeqnarray}
%\sqrt{(\Delta m)^2}
 {{E_{p} =  2.2 {\left(
\frac{4eV} {\sqrt{{{m_{\nu}}^2+{p_{\nu}^2}}}} \right)} \cdot
10^{19} \,eV.} \nonumber}
\end{subeqnarray}

   We usually may consider cosmological relic neutrinos in Standard Model
   at non relativistic regime
   neglecting ${p_{\nu}} $ term.  However, at lightest mass values
   the momentum may be comparable to the relic mass; moreover
   the spectra may reflect unexpected relic neutrino black bodies or gray body
   at energies much above the neutrino mass. Indeed there may be
   exist, within or beyond Standard Cosmology,  a relic neutrino component due to stellar,Super Nova,GRBs,AGN activities
   red-shifted  into a present KeV-eV relic neutrino grey-body  energy spectra.
     Therefore  it is worth-full to keep the most general
      mass and momentum term in the relic neutrino energy.

As we noticed above, relic neutrino mass above a few eVs in  HDM
halo are not consistent with Z peak; higher energies interactions
ruled by WW,\cite{Enq},\cite{FarSal99} ZZ cross-sections  may
nevertheless solve the GZK cut-off . In this regime there will be
also possible to produce by virtual W exchange, t-channel, $UHE$
lepton pairs, by $\nu_i \bar{\nu}_j\rightarrow \l_i
\bar{\l}_j$, leading to additional electro-magnetic showers injection.\\
 The hadronic tail of the Z or $W^+ W^-$ cascade is
 the source of final  nucleons $p,\bar{p}, n, \bar{n}$ able to explain UHECR events observed by
Fly's Eye and AGASA \cite{AGASA} and other detectors. The same
$\nu \bar{\nu}_r$ interactions are source of Z and W that decay in
rich shower ramification. The electro-magnetic showering will be
discussed in detail else-where \cite{Fargion2001b}. The average
energy deposition for both gauge bosons among the secondary
particles is summarized in Table 1

\begin{table}[h]
\begin{center}
\begin{tabular}{|c|c|c|c|}
\hline
   & Z & $W^+ W^-$ & t-channel\\ \hline
  $\nu$ & 58 \% & 55 \% & 47 \% \\ \hline
    $\gamma$
& 21 \% & 21 \% & 4 \% \\ \hline
    $e^+ e^-$ & 16 \% & 19 \% & 49 \% \\ \hline
  $p$ & 5 \% & 5 \% & - \\ \hline
\end{tabular}
\end{center}

\caption{Total Energy percentage distribution  into neutrino,
gamma, electron pairs particles (from Z and $WW,ZZ$  as well as
t-channel W decay), before energy losses. These UHE photons are
mainly relics of neutral pions. Most of the $\gamma$ radiation
will be degraded around PeV energies by $\gamma \gamma$ pair
production with cosmic 2.75 K BBR, or with cosmic radio
background. The electron pairs instead, are mainly relics of
charged pions and will rapidly lose energies into synchrotron
radiation   } \label{1}
\end{table}

 Although protons (or anti-protons, as well as neutron and anti-neutrons)
  are the most favorite candidate in
order to explain the highest energy air shower observed, one
doesn't have to neglect the signature of final electrons and
photons. In fact electron (positron) interactions with the
galactic magnetic field or soft radiative backgrounds may lead to
gamma cascades and it may determine gamma signals from EeV, to MeV
energies related to the same UHECR shower event .\\
Gamma photons at energies $E_{\gamma} \simeq 10^{20}$ - $10^{19}
\,eV$ may freely propagate through galactic or local halo scales
(hundreds of kpc to few Mpc) and could contribute to the extreme
edges of cosmic ray spectrum \cite{Yoshida}\cite{Fargion2001b}. \\
The ratio of the final energy flux of nucleons near the Z peak
resonance, $\Phi_p$ over the corresponding electro-magnetic
energy flux $\Phi_{em}$ ratio is, as in tab.1 $e^+ e^-,\gamma$
entrance, nearly $\sim \frac{1}{8}$.  Moreover if one considers
at higher $E_{\nu}$ energies, the opening of WW, ZZ channels and
the six pairs $\nu_e \bar{\nu_{\mu}}$, \, $\nu_{\mu}
\bar{\nu_{\tau}}$, \, $\nu_e \bar{\nu_{\tau}}$ (and their
anti-particle pairs) t-channel interactions leading to  highest
energy leptons, with no nucleonic relics (as $p, \bar{p}$), this
additional injection favors the electro-magnetic flux $\Phi_{em}$
over the corresponding nuclear one $\Phi_p$ by a factor $\sim
1.6$ leading to $\frac{\Phi_p}{\Phi_{em}} \sim \frac{1}{13}$.
This ratio is valid at $WW,ZZ$ masses because the overall cross
section variablility is energy dependent. At center of mass
energies above these values, the $\frac{\Phi_p}{\Phi_{em}}$
decreases more because the dominant role of t-channel (Fig1). We
shall focus here  on Z, and WW,ZZ channels showering in hadrons
while their main consequent electro-magnetic showering
 will be discussed  elsewhere \cite{Fargion2001b}. \\

Extragalactic neutrino cosmic rays are free to move on cosmic
distances up our galactic halo without constraint on their mean
free path, because the interaction length with cosmic background
neutrinos is greater than the actual Hubble distance . A Hot Dark
Matter galactic or local group halo model with relic light
neutrinos (primarily the heaviest $\nu_{\tau}$ or $ \nu_{\mu} $)
\cite{FarSal99}, acts as a target for the high energy neutrino
beams. The relic number density and the halo size are large
enough to allow the $\nu \nu_{relic}$ interaction . As a
consequence high energy particle showers are produced in the
galactic or local group halo, overcoming the GZK cut-off
\cite{FarSal99}.  There is an upper bound density clustering for
very light Dirac fermions due to the maximal Fermi degenerancy
whose adimensional density contrast is $\delta\rho \propto
m_{\nu}^3$, \cite {Fargion 83},\cite {FarSal99}, while the
neutrino free-streaming halo grows only as $\propto
m_{\nu}^{-1}$. Therefore the overall interaction probability
grows $ \propto m_{\nu}^{2} $, favoring heavier non relativistic
(eVs) neutrino masses. Nevertheless the same lightest relic
neutrinos may share higher Local Group velocities (thousands
$\frac{Km}{s}$) or even nearly relativistic speeds and it may
therefore compensate the common bound:

\begin{equation}
 n_{\nu_{i}}=10^{3}\left( \frac{n_{\nu_{i}}}{54cm^{-3}}\right)
\left( \frac{m_{i}}{0.1eV}\right)  ^{3}\left(
\frac{v_{\nu_{i}}}{2000\frac{Km}{s} }\right)  ^{3}
\end{equation}

%\begin{equation} { {n_{\nu_{\i}}}= {10^3 {\left(
%\frac{n_{\nu_i}}{54\,{\rm cm}^{-3}}\right)} \;
%\left(\frac{m_{\nu_i}}{{\rm 0.1eV}}\right)^3 \left( \frac{\
%v_{\nu_i}}{2000{\rm km/s}}\right)^3\,\right}} \end{equation}

%\\

 From the cross section side there are three main interaction processes that
 have to be considered  leading to nucleons in the
of EHE and relic neutrinos scattering.\\

{\bf channel 1.} $\;$ The $\nu \nu_r\rightarrow Z \rightarrow \, $
 annihilation at the Z resonance \\

{\bf channel 2.} $\nu_{\mu} \bar{\nu_{\mu}} \rightarrow W^+ W^-$
or $\nu_{\mu} \bar{\nu_{\mu}} \rightarrow Z Z$ leading to hadrons, electrons, photons,
through W and Z decay. \\

{\bf channel 3.} The $\nu_e$ - $\bar{\nu_{\mu}}$, $\nu_e$ -
$\bar{\nu_{\tau}}$, $\nu_{\mu}$ - $\bar{\nu_{\tau}}$ and hermite
conjugate interactions of different flavor neutrinos mediated in
the $t$ - channel by the W exchange (i.e. $\nu_{\mu}
\bar{\nu_{\tau_r}} \rightarrow \mu^- \tau^+ $). These reactions
are sources of prompt and secondary UHE electrons as well as
photons resulting by hadronic
$\tau$ decay.\\

\subsection{The process $\nu_{\tau} \bar{\nu_{\tau}} \rightarrow Z $}

 The interaction of neutrinos of the
same flavor can occur via a Z exchange in the $s$-channel
($\nu_i\bar{\nu}_{i_r}$ and charge conjugated). The cross section
for hadron production in $\nu_i\bar{\nu}_i\rightarrow
Z^*\rightarrow hadrons$ is
\begin{equation}
\sigma_Z(s)=\frac{8\pi s}{M_Z^2}\frac{\Gamma(Z^o\rightarrow
invis.) \Gamma(Z^o\rightarrow hadr.)}{(s-M_Z^2)^2+M_Z^2 \Gamma_Z^2}
\end{equation}
where $\Gamma(Z^o\rightarrow invis.)\simeq 0.5~GeV$,
$\Gamma(Z^o\rightarrow hadr.)\simeq 1.74~GeV$ and $\Gamma_Z\simeq
2.49~GeV$ are respectively the experimental Z width into invisible
products, the Z width into hadrons and the Z full width \cite{pdg}
. The averaged cross section peak reaches the value ($< \sigma_Z >
= 4.2 \cdot 10^{-32} \, cm^2$).  We assumed here for a more
general case (non relativistic and nearly relativistic relic
neutrinos)  that the averaged cross section has to be extended
over an energy window
comparable to half the center of mass energy. The consequent effective
averaged cross-section is described in Fig.1 as a truncated hill curve.\\
A $\nu\nu_r$ interaction mediated in the $s$-channel by the Z
exchange, shows a peculiar peak in the cross section due to the
resonant Z production at $s= M_Z^2$. However, this occurs for a
very narrow and fine-tuned windows of  arrival neutrino energies
${\nu}_{\i}$  (and of the corresponding target neutrino masses
and momentum $ \bar{\nu}_ {\i}$ ):

\begin{equation}
{E_{\nu_{\i}}} =  {\left( \frac{4eV} {\sqrt {{{m_{\nu_
{\i}}}^2+{p_{\nu_ {\i}}}^2}}} \right)} \cdot 10^{21} \,eV.
\end{equation}

So in this mechanism the energy of the EHE neutrino cosmic ray is
related to the mass of the relic neutrinos, and for an initial
neutrino energy fixed at $E_{\nu} \simeq 10^{22} \, eV$, the Z
resonance requires a mass for the heavier neutral lepton around
$m_{\nu} \simeq 0.4 \, eV$. Apart from this narrow
 resonance peak at $\sqrt{s}= M_Z$, the
asymptotic behaviour of the cross section is proportional to $1/s$ for $s\gg M_Z^2$.
\\

The $\nu \bar{\nu} \rightarrow Z \rightarrow hadrons$ reactions
have been proposed by \cite {FarSal97} \cite{Weiler}
\cite{Yoshida} with a neutrino clustering on Supercluster,
cluster, Local Group, and galactic halo scale within the few tens
of Mpc limit fixed by the GZK cut-off. Due to the enhanced
annihilation cross-section in the Z pole, the probability of a
neutrino collision is reasonable even for a low neutrino density
contrast $\delta \rho_{\nu} / \rho_{\nu} \geq 10^3$. The
potential wells of such structures might enhance the neutrino
local density with an efficiency at comparable with observed
baryonic clustering discussed above.

%%%%%%%%%%%%%%%   Figure 1 Ex 01  %%%%%%%%%%%%%%%%%   FFFFFFFFFFFFFFFFFFFFFFFFFFFFFFFFFFFFFF
\begin{figure}[h]
\begin{center}
 \includegraphics[width=0.9\textwidth, height=7.0cm] {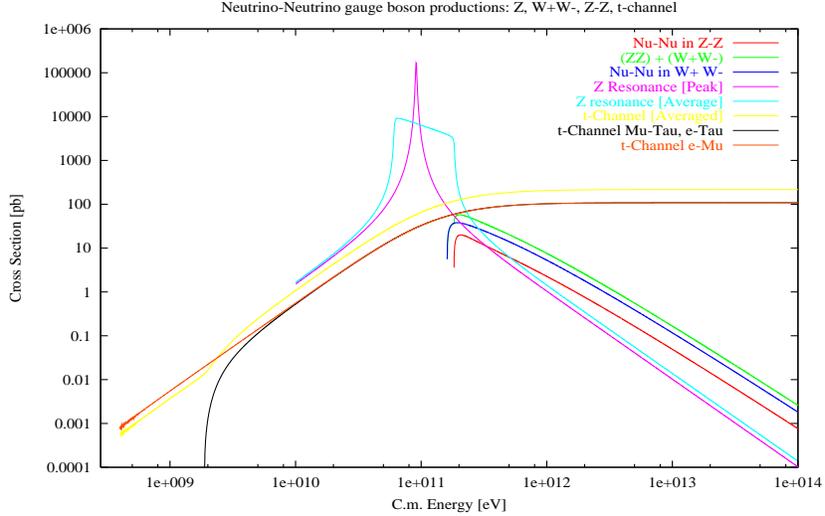}
\end{center}
  \caption{The  $\nu \bar{\nu} \rightarrow  Z,W^+ W^-,ZZ
  ,T$-channel,  cross sections as a function of the center of mass energy in $\nu \nu$.
   These cross-sections are estimated also in average (Z) as well for each possible
   t-channel lepton pairs. The averaged t-channel averaged the multiplicity of flavours
    pairs ${\nu}_{\i}$, $ \bar{\nu}_{\j}$ respect to neutrino
    pair annihilations into Z neutral boson. } \label{fig:boxed_graphic 1}
\end{figure}
%%%%%%%%%%%%%%%   Figure 1 END   %%%%%%%%%%%%%%%%%   FFFFFFFFFFFFFFFFFFFFFFFFFFFFFFFFFFFFFF

\subsection{The processes  $\nu_{\tau}
\bar{\nu_{\tau}} \rightarrow W^+ W^-$ and $\nu_{\tau}
\bar{\nu_{\tau}} \rightarrow Z Z $}

%The Z resonant neutrino annihilation is about three orders of
%magnitude higher than W pair channel , but it occurs for a very narrow and
%fine-tuned window of energy

The reactions $\nu_{\tau} \bar{\nu_{\tau}} \rightarrow W^+ W^-$,%
 $\nu_{\mu} \bar{\nu_{\mu}} \rightarrow W^+ W^- $,%
 $\nu_{e}  \bar{\nu_{e}} \rightarrow W^+ W^-$, %
 that occurs through the exchange of a Z boson (s channel)) \cite{Enq}, has been
previously introduced  \cite{FarSal99} in order to explain UHECR
as the Fly's Eye event at 320 Eev detected in 1991 and last AGASA
data. The cross section is given by \cite{FarSal99}

\begin{equation}
\sigma_{WW}(s)=\sigma_{asym}\frac{\beta_W}{2s}\frac{1}{(s-M_Z^2)}\left\{4
L(s) \cdot C(s)+D(s)\right\}
\end{equation}
where $\beta_W=(1-4 M_W^2/s)^{1/2}$,
$\sigma_{asym}=\frac{\pi\alpha^2}{2\sin^4\theta_W
M_W^2}\simeq~108.5~pb$, and the functions $L(s)$, $C(s)$, $D(s)$
are defined as
\begin{displaymath}
L(s)=\frac{M_W^2}{2\beta_W s}\ln\Big(\frac{s+\beta_W s-2 M_W^2}{s-
\beta_W s-2 M_W^2}\Big)
\end{displaymath}
\begin{equation}
C(s)=s^2+s(2 M_W^2-M_Z^2)+2 M_W^2(M_Z^2+M_W^2)
\end{equation}
\begin{displaymath}
D(s)=\frac{1}{12 M_W^2 (s-M_Z^2)}\times \Big[ s^2(M_Z^4-60
M_W^4-4 M_Z^2 M_W^2) +
\end{displaymath}
\begin{displaymath}
 +20 M_Z^2 M_W^2 s ( M_Z^2+2 M_W^2)-48 M_Z^2
M_W^4(M_Z^2 + M_W^2) \Big]~.
\end{displaymath}\\

This result should be compared with the additional new  ZZ interaction channel
 considered for the first time here:\\

\begin{equation}\label{4}
  \sigma_{ZZ} = \frac{G^2M^2_Z}{4 \pi} y \frac{(1 + \frac{y^2}{4})}{(1 - \frac{y}{2})}
  \left \{  \ln \left[ \frac{2}{y} (1 - \frac{y}{2} + \sqrt{1 - y}) \right
  ]-\sqrt{1 - y} \right \}
\end{equation}

where $y = \frac{4M^2_Z}{s}$ and $\frac{G^2M^2_Z}{4 \pi} = 35.2
\,pb$.\\

Their values are plotted in Fig.1.
 The asymptotic behaviour of
these cross section is proportional to
$\sim(\frac{M_W^2}{s})\ln{(\frac{s}{M_W^2})}$ for $s\gg M_Z^2$.\\
The nucleon arising from WW and ZZ hadronic decay could provide a
reasonable solution to the 320 Eev event puzzle. We'll assume
that the fraction of pions and nucleons related to the total
number of particles from the W boson decay is the almost the same
of Z boson. So W hadronic decay ($P \sim 0.68$) leads on average
to about 37 particles, where $<n_{\pi^0}> \sim 9.19$,
$<n_{\pi^{\pm}} > \sim 17$, and $<n_{p,\bar{p}, n, \bar{n}}> \sim
2.7$. In addition we have to expect by the subsequent decays of
$\pi$'s (charged and neutral), kaons and resonances ($\rho$,
$\omega$, $\eta$) produced, a flux of secondary UHE photons and
electrons.\\ As we already pointed out, the particles resulting
from the decay are mostly prompt pions. The others are particles
whose final decay likely leads to charged and neutral pions as
well. As a consequence the electrons and photons come from prompt
pion decay. \\ On average it results \cite{pdg} that the energy
in the bosons decay is not uniformly distributed among the
particles, so that proton energy is about three times that of the
direct pions. Each charged pion will give an electron (or
positron) and three neutrinos, that will have less than one per
cent of the initial W boson energy, while each $\pi^0$ decays in
two photons, each with 1 per cent of the initial W energy . In
the Table1 below we show all the channels leading from single Z,W
and Z pairs as well as t-channel in nuclear and electro-magnetic
components.
Their energies and corresponding fluence are summirized in Table 2.\\
%At the same time a pure W leptonic decay $W \rightarrow l \nu$
%could occur for each flavor with a probability $P \sim 0.11$
%(see Table 4).\\

\subsection{The process  $\nu_{i} \nu_{j} \rightarrow l_i l_j$: the t-channel}

The processes $\nu_{i} \nu_{j} \rightarrow l_i l_j$ (like
$\nu_{\mu} \nu_{\tau} \rightarrow \mu \tau $ for example)
 \footnote{We could
consider as well the reactions $\nu_{e} \bar{\nu_{\tau_r}}
\rightarrow e^- \tau^+$, $\nu_{e} \bar{\nu_{\mu_r}} \rightarrow
e^- \mu^+$ and $\nu_{e} \bar{\nu_{e_r}} \rightarrow e^- e^+$,
changing the target or the high energy neutrino. Therefore there
are  2 times more target than for Z, WW, ZZ channels.}
 occur through the W boson exchange in the t-channel.
The cross-section has been derived in \cite{FarSal99}, while the
energy threshold depends on the mass of the heavier lepton
produced, $E_{\nu_{th}} = 7.2 \cdot 10^{19}(m_{\nu} / 0.4 \,
eV)^{-1} (m_{\tau} / m_{\tau , \mu , e})$, with the term
$(m_{\tau} / m_{\tau ,\mu , e})$ including the different
thresholds in all the possible interactions: $\nu_{\tau}
\nu_{\mu}$ (or $\nu_{\tau} \nu_e)$ , $\nu_{\mu} \nu_{e}$, and
$\nu_{e} \nu_{e}$. In the ultrarelativistic limit ($s \simeq
2E_{\nu} m_{\nu_r} \gg M^2_W$ where $\nu_r$ refers to relic
clustered neutrinos)  the cross-section tends to the
asymptotic value $\sigma_{\nu \bar{\nu_r}} \simeq 108.5 \,pb$.\\

\begin{equation}
 \sigma_W(s)= \sigma_{asym}\frac{A(s)}{s}\left\{
1+\frac{M_W^2}{s}\Bigg[
2-\frac{s+B(s)}{A(s)}\ln\Bigg(\frac{B(s)+A(s)}{B(s)-
A(s)}\Bigg)\Bigg]\right\}
\end{equation}
where $\sqrt{s}$ is the center of mass energy, the functions
A(s), B(s) are defined as
\begin{equation}
A(s)=\sqrt{[s-(m_\tau+m_\mu)^2][s-(m_\tau-m_\mu)^2]}~;~
B(s)=s+2M_W^2-m_\tau^2-m_\mu^2
\end{equation}
and
\begin{equation}
\sigma_{asym}=\frac{\pi\alpha^2}{2\sin^4\theta_W
M_W^2}\simeq~108.5~pb
\end{equation}
where $\alpha$ is the fine structure constant and $\theta_W$ the
Weinberg angle. $\sigma_{asym}$ is the asymptotic behaviour of
the cross section in the ultrarelativistic limit
\begin{equation}
s\simeq 2 E_\nu m_\nu= 2\cdot 10^{23} (E_\nu /10^{22}~eV)(m_\nu /
10~eV)~eV^2 \gg M_W^2~~~~~.
\end{equation}

This interactions,as noted in Table.1 are leading to
electro-magnetic showers and are not offering any nuclear
secondary. Their astrophysical role will be discussed  elsewhere
\cite {Fargion2001b}.

\section{The prediction of the UHE particles spectra from W and Z decay}

Let us examine the destiny of UHE primary particles (nucleons,
electrons and photons) ($E_e \lesssim 10^{21}\,eV$) produced
after hadronic or leptonic W decay. As we already noticed in the
introduction, we'll assume that the nucleons, electrons and
photons spectra (coming from W or Z decay) after $\nu \nu$
scattering in the halo, follow a power law that in the center of
mass system is $\frac{dN^*}{dE^* dt^*} \simeq E^{* - \alpha}$
where $\alpha \sim 1.5$. This assumption is based on detailed
Monte Carlo simulation of a heavy fourth generation neutrino
annihilations \cite{Konoplich} \cite{Konoplich2}and with the
model of quark - hadron fragmentation spectrum suggested by Hill
\cite{Hill}.

In order to determine the shape of the particle spectrum in the
laboratory frame, we have to introduce the Lorentz relativistic
transformations from the center of mass system to the laboratory
system. \\
 The number of particles is clearly a relativistic invariant $dN_{lab} = dN^*$,
while the relation between the two time intervals is $dt_{lab} =
\gamma dt^*$, the energy changes like $ \epsilon_{lab} = \gamma
\epsilon^* (1 + \beta \cos \theta^*) = \epsilon^* \gamma^{-1}(1 -
\beta \cos \theta)^{-1}$, and finally the solid angle in the
laboratory frame of reference becomes $d\Omega_{lab} =\gamma^{2}
d\Omega^*  (1 - \beta \cos \theta )^2$. Substituting these
relations one obtains

%\begin{subeqnarray}
%\left(\frac{dN}{d\epsilon dt d\Omega} \right)_{lab} =
%\frac{dN_{*}}{d\epsilon_{*} dt_{*} d\Omega_{*}} \gamma^{-2}
% (1 - \beta \cos \theta)^{-1} =
% \frac{\epsilon^{-\alpha}_{*} \; \gamma^{-2}} {4 \pi}} \cdot (1 - \beta \cos
% \theta)^{-1} \nonumber \\
%\left( \frac{dN}{d\epsilon dt d\Omega} \right)_{lab} =
% \frac{\epsilon^{-\alpha} \; \gamma^{-\alpha-2}} {4 \pi}} (1 - \beta \cos \theta)^{-\alpha-1}\setcounter{eqsubcnt}{0}
%\end{subeqnarray}

\begin{subeqnarray}
{{ \left(\frac{dN}{d\epsilon dt d\Omega} \right)_{lab} =
\frac{dN_{*}}{d\epsilon_{*} dt_{*} d\Omega_{*}} \gamma^{-2}
 (1 - \beta \cos \theta)^{-1} =
 \frac{\epsilon^{-\alpha}_{*} \; \gamma^{-2}} {4 \pi}} \cdot (1 - \beta \cos
 \theta)^{-1} \nonumber}
\end{subeqnarray}

\begin{equation}
{{\left( \frac{dN}{d\epsilon dt d\Omega} \right)_{lab} =
 \frac{\epsilon^{-\alpha} \; \gamma^{-\alpha-2}} {4 \pi}} (1 - \beta \cos \theta)^{-\alpha-1}\setcounter{eqsubcnt}{0}
 }
\end{equation}

and integrating on $\theta$ (omitting the lab notation) one loses
the spectrum dependence on the angle.

%\begin{equation}
%\left( \frac{dN}{d\epsilon dt d\Omega} \right)_{lab} \propto
%\epsilon^{-\alpha} \gamma_{Z (W)}^{ \alpha} \sim \epsilon^{-
%\frac{\alpha}{2}} \sim \epsilon^{- 0.75}.
%\end{equation}

The consequent fluence derived by the solid angle integral is:
% Long version
%\begin{equation}
% \frac{dN}{d\epsilon dt} \epsilon^{2}=
% \frac{\epsilon^{-\alpha+2} \; \gamma^{\alpha-2}} {2 \beta \alpha}}
% [(1 + \beta)^{\alpha} -
%  \frac{1} {[(1 + \beta)\gamma^2]^{\alpha}}}] \simeq
% \frac{\epsilon^{-\alpha+2} \; \gamma^{\alpha-2}} {\alpha}}
%\end{equation}

\begin{equation}
{ \frac{dN}{d\epsilon dt} \epsilon^{2}=
 \frac{\epsilon^{-\alpha+2} \; \gamma^{\alpha-2}} {2 \beta \alpha}
 [(1 + \beta)^{\alpha} - (1 - \beta)^{\alpha}] \simeq
 \frac{2^{\alpha-1}\epsilon^{-\alpha+2} \; \gamma^{\alpha-2}} {\alpha}}
\end{equation}

There are to extreme case to be considered: the case where the
interaction occur at Z peak resonance and therefore the center of
mass Lorents factor $\gamma$ is "frozen" at a given value (eq.1)
and the case (WW,ZZ pair channel) where all energies are
allowable  and $\gamma$ is proportional to $\epsilon^{1/2}$.
%; the latter case will be discussed in detail elsewhere.
 Here we focus only on Z peak resonance. The consequent fluence spectra
 $\frac{dN}{d\epsilon dt}\epsilon^{2}$, as above, is proportional to $\epsilon^{-\alpha +2}$. Because $\alpha$ is
nearly $1.5$ all the consequent secondary particles will also show
a spectra proportional to $\epsilon^{1/2}$ following a normalized
energies shown in Tab.2, as shown in Fig.(2-6). In the latter
case (WW,ZZ pair channel), the relativistic boost reflects on the
spectrum of the secondary particles, and the spectra power law
becomes $\propto \epsilon^{\alpha/2 +1}=\epsilon^{0.25}$. These
channels will be studied in details elsewhere. In Fig. 1 we show
the spectrum of protons, photons and electrons coming from Z
hadronic and leptonic decay assuming a nominal primary CR energy
flux $\sim 20 eV s^{-1} sr^{-1} cm^{-2}$, due to the total $\nu
\bar{\nu}$ scattering at GZK energies as shown in figures 2-6.
Let us remind that we assume an interaction probability of $\sim
1 \%$ and a corresponding UHE incoming neutrino energy $\sim 2000
eV s^{-1} sr^{-1} cm^{-2}$ near but below present $UHE$ neutrino
flux bound.

\vspace{0.3cm}

\begin{table}[h]
\begin{center}
\begin{tabular}{|c|c|c|} \hline
  % after \\ : \hline or \cline{col1-col2} \cline{col3-col4} ...
  $Z_{decay}$ & E (eV) & $\frac{dN}{dE}E^2$ (eV) \\ \hline
  p & $2.2 \cdot 10^{20}$ & 1.2 \\ \hline
  $\gamma$ & $9.5 \cdot 10^{19}$ & 4.25 \\ \hline
  $e_{\pi}$ & $5 \cdot 10^{19}$ & 2.4 \\ \hline
   $e_{prompt}$ & $5 \cdot 10^{21}$ & 0.66 \\ \hline
    $e_{\mu}$ & $1.66 \cdot 10^{21}$ & 0.23 \\ \hline
     $e_{\tau}$ & $1.66 \cdot 10^{21}$ & 0.12 \\ \hline
\end{tabular}
\end{center}
\caption{Energy peak and Energy Fluence for different decay
channels as described in the text.} \label{2}
\end{table}

\vspace{0.3cm}

\begin{figure}[h]
\begin{center}
 \includegraphics[width=0.9\textwidth, height=7.0cm] {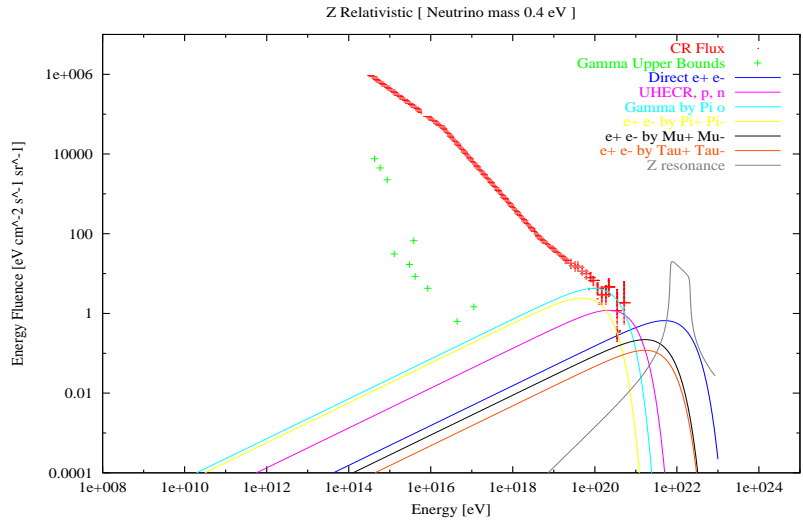}
\end{center}
  \caption{Energy Fluence derived by $\nu \bar{\nu} \rightarrow Z$ and its showering into
  different channels: direct electron pairs UHECR nucleons $n$ $p$ and anti-nucleons, $\gamma$ by $\pi^0$ decay,
  electron pair by $\pi^+ \pi^-$ decay, electron pairs by direct muon and tau decays as labeled in figure.
  The relic neutrino mass has been assumed to be fine tuned to explain GZK UHECR tail:
  $m_{\nu}=0.4 eV$. The "Z resonance ghost" curve, derived from averaged cross-section
  in Fig.1, shows the averaged $Z$ resonant cross-section peaked
  at $E_{\nu}=10^{22} eV$. Each channel shower has been normalized following table 2.}
\label{fig:boxed_graphic 2}
\end{figure}
%%%%%%%%%%%%%%%   Figure 2 END   %%%%%%%%%%%%%%%%%   FFFFFFFFFFFFFFFFFFFFFFFFFFFFFFFFFFFFFF

%%%%%%%%%%%%%%%   Figure 3       %%%%%%%%%%%%%%%%%   FFFFFFFFFFFFFFFFFFFFFFFFFFFFFFFFFFFFFF
\begin{figure}[h]
\begin{center}
 \includegraphics[width=0.9\textwidth, height=7.0cm] {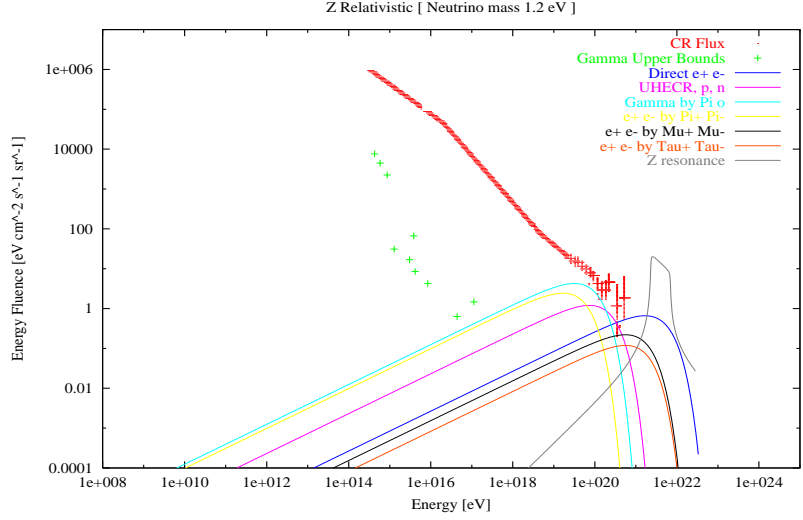}
\end{center}
  \caption{Energy Fluence derived by $\nu \bar{\nu} \rightarrow Z$ and its showering into
  different channels as in previous Figure 2: direct electron pairs UHECR nucleons $n$ $p$, $\gamma$ by $\pi^0$ decay,
  electron pair by $\pi^+ \pi^-$ decay, electron pairs by direct muon and tau decays as labeled in figure.
  In the present case the relic neutrino mass has been assumed to be fine tuned to explain GZK UHECR tail:
  $m_{\nu}=1.2 eV$ with the same UHE incoming neutrino fluence of previous figure. The "Z resonance" curve shows the averaged $Z$ resonant cross-section peaked
  at $E_{\nu}=3.33\cdot10^{21} eV$.Each channel shower has been normalized in analogy to table 2.}
  \label{fig:boxed_graphic 3}
\end{figure}
%%%%%%%%%%%%%%%   Figure 3 END   %%%%%%%%%%%%%%%%%   FFFFFFFFFFFFFFFFFFFFFFFFFFFFFFFFFFFFFF

%%%%%%%%%%%%%%%   Figure 4       %%%%%%%%%%%%%%%%%   FFFFFFFFFFFFFFFFFFFFFFFFFFFFFFFFFFFFFF
%\begin{figure}[h]
%\begin{center}
% \includegraphics[width=0.9\textwidth, height=7.0cm] {nu36.eps}
%\end{center}
%  \caption{Energy Fluence derived by $\nu \bar{\nu} \rightarrow Z$ and its showering into
%  different channels: direct electron pairs UHECR nucleons $n$ $p$, $\gamma$ by $\pi^0$ decay,
%  electron pair by $\pi^+ \pi^-$ decay, electron pairs by direct muon and tau decays as labeled in figure.
%  In the present case the relic neutrino mass has been assumed to be fine tuned to explain GZK UHECR tail:
%  $m_{\nu}=3.6 eV$ with an increased UHE incoming neutrino fluence by a factor 3.
%   The "Z resonance" curve shows the averaged $Z$ resonant cross-section peaked
%  at $E_{\nu}=10^{21} eV$. Each channel shower has been normalized in analogy to table 2.}
%\label{fig:boxed_graphic 4}
%\end{figure}
%%%%%%%%%%%%%%%   Figure 4 END   %%%%%%%%%%%%%%%%%   FFFFFFFFFFFFFFFFFFFFFFFFFFFFFFFFFFFFFF

%%%%%%%%%%%%%%%   Figure 4       %%%%%%%%%%%%%%%%%   FFFFFFFFFFFFFFFFFFFFFFFFFFFFFFFFFFFFFF
\begin{figure}[h]
\begin{center}
 \includegraphics[width=0.9\textwidth, height=7.0cm] {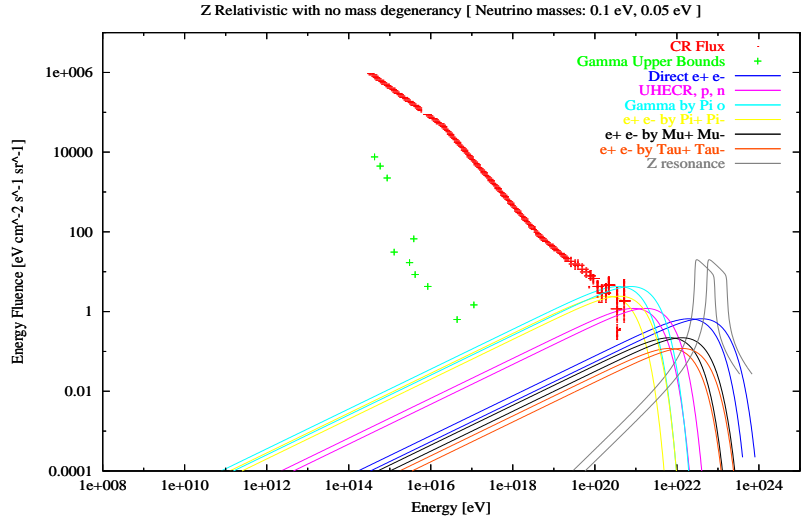}
\end{center}
  \caption{Energy Fluence derived by $\nu \bar{\nu} \rightarrow Z$ and its showering into
  different channels: direct electron pairs UHECR nucleons $n$ $p$, $\gamma$ by $\pi^0$ decay,
  electron pair by $\pi^+ \pi^-$ decay, electron pairs by direct muon and tau decays as labeled in figure.
  In the present case the relic neutrino masses have been assumed with no degenerancy.
  The their values have been fine tuned to explain GZK UHECR tail:
   $m_{\nu_1}=0.1 eV$ and $m_{\nu_2}=0.05 eV$. No relic neutrino
   density difference has been assumed.
   The incoming UHE neutrino fluence has been increased
   by a factor 2 respect previous Fig.2-3. The "Z resonance" curve shows the averaged $Z$ resonant cross-section peaked
  at $E_{\nu_1}=4\cdot10^{22} eV$ and $E_{\nu_2}=8\cdot10^{22} eV$. Each channel shower has been normalized in analogy to table 2.}
\label{fig:boxed_graphic 4} % ma e' la quattro....
\end{figure}
%%%%%%%%%%%%%%%   Figure 4 END   %%%%%%%%%%%%%%%%%   FFFFFFFFFFFFFFFFFFFFFFFFFFFFFFFFFFFFFF

%%%%%%%%%%%%%%%   Figure 5       %%%%%%%%%%%%%%%%%   FFFFFFFFFFFFFFFFFFFFFFFFFFFFFFFFFFFFFF
\begin{figure}[h]
\begin{center}
 \includegraphics[width=0.9\textwidth, height=7.0cm] {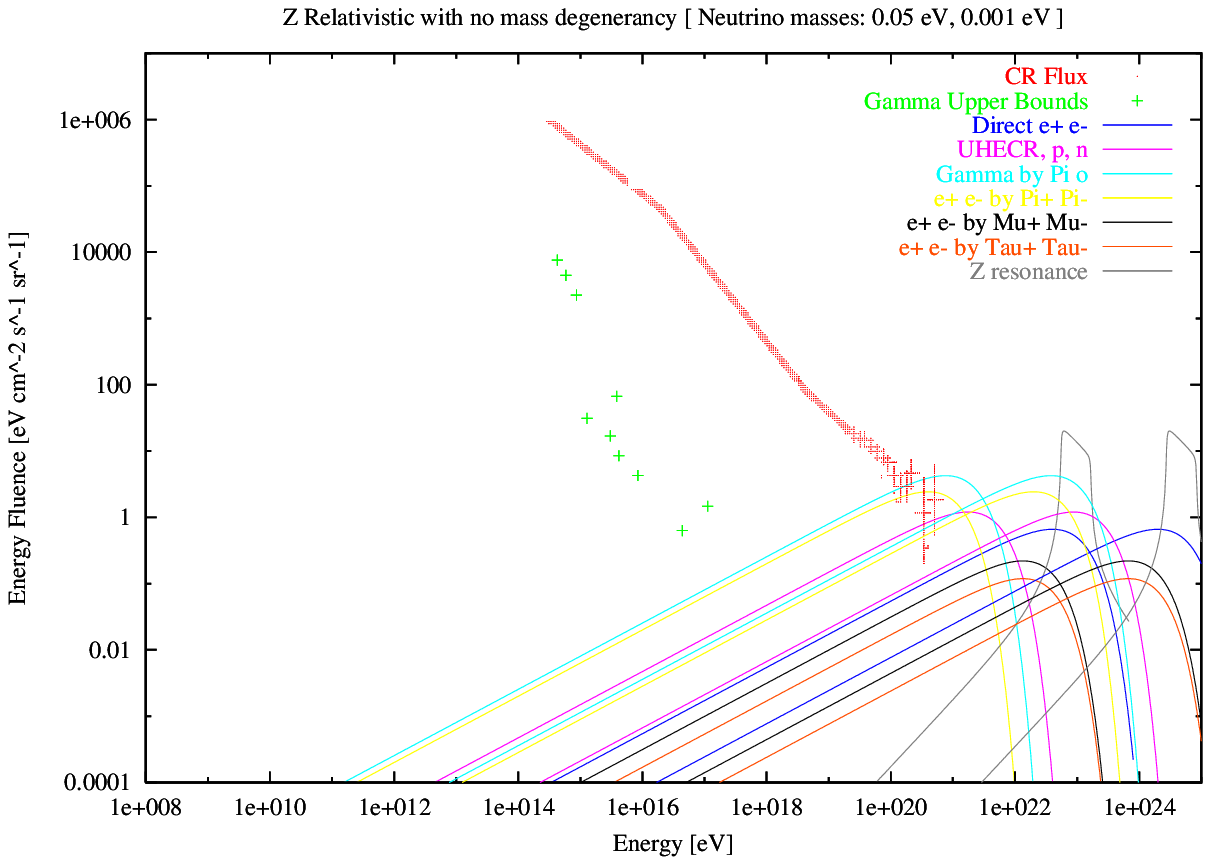}
\end{center}
\caption{Energy Fluence derived by $\nu \bar{\nu} \rightarrow Z$
and its showering into
  different channels  as above.
  In the present extreme case the relic neutrino masses have been assumed with wide mass differences
  just compatible both with Super-Kamiokande and relic $2 K^{o}$ Temperature .
  The their values have been fine tuned to explain observed GZK- UHECR tail:
   $m_{\nu_1}=0.05eV$ and $m_{\nu_2}=0.001 eV$. No relic neutrino
   density difference between the two masses  has been assumed,
   contrary to bound in eq.3.
   The incoming UHE neutrino fluence has been increased
   by a factor 2 respect previous Fig.2-3. The "Z resonance" curve
    shows the averaged $Z$ resonant cross-section peaked
  at $E_{\nu_1}=8\cdot10^{22} eV$ and $E_{\nu_2}=4\cdot10^{24} eV$, just
  near Grand Unification energies. Each channel shower has been normalized in analogy to table 2.}
\label{fig:boxed_graphic 5}
\end{figure}
%%%%%%%%%%%%%%%   Figure 5 END   %%%%%%%%%%%%%%%%%   FFFFFFFFFFFFFFFFFFFFFFFFFFFFFFFFFFFFFF

%%%%%%%%%%%%%%%   Figure 6       %%%%%%%%%%%%%%%%%   FFFFFFFFFFFFFFFFFFFFFFFFFFFFFFFFFFFFFF
\begin{figure}[h]
\begin{center}
 \includegraphics[width=0.9\textwidth, height=7.0cm] {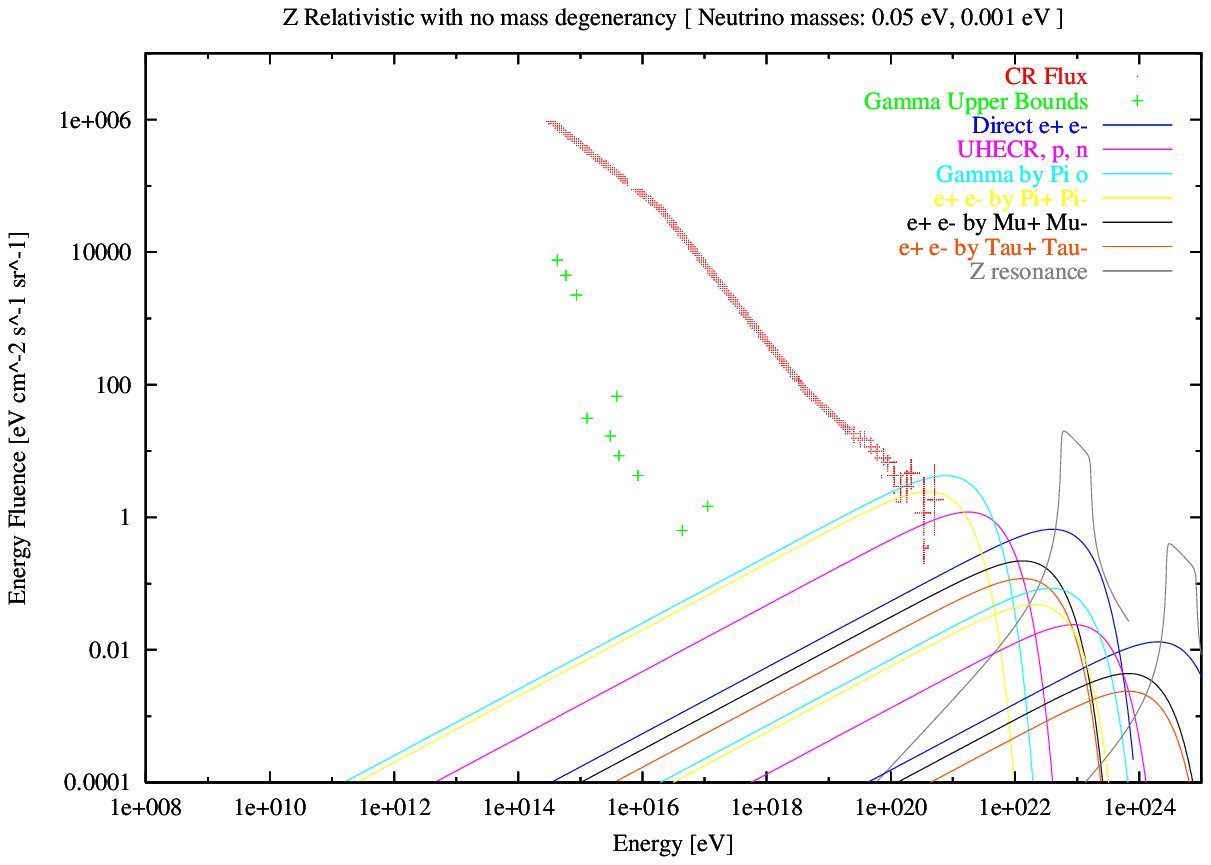}
\end{center}
\caption{Energy Fluence derived by $\nu \bar{\nu} \rightarrow Z$
and its showering into
  different channels  as above.
  In the present extreme case the relic neutrino masses have been assumed with wide mass differences
  just compatible both with Super-Kamiokande and relic $2 K^{o}$ Temperature .
  The their values have been fine tuned to explain observed GZK- UHECR tail:
   $m_{\nu_1}=0.05eV$ and $m_{\nu_2}=0.001 eV$. A neutrino
   density difference between the two masses  has been
   assumed,considering the lightest $m_{\nu_2}=0.001 eV$ neutrino
   at relativistic regime, consistent to bound in eq.3.
   The incoming UHE neutrino fluence has been assumed growing
   linearly \cite{Yoshida}  with energy. Its value is increased
   by a factor 2 and 20  at
   $E_{\nu_1}=8\cdot10^{22} eV$ and $E_{\nu_2}=4\cdot10^{24} eV$
   respect the previous ones Fig.2-3. The "Z resonance" curve
    shows its averaged $Z$ resonant "ghost" cross-section peaked
  at $E_{\nu_1}=2\cdot10^{23} eV$ and $E_{\nu_2}=4\cdot10^{24} eV$, just
  near Grand Unification energies.
   Each channel shower has been normalized in analogy to table 2.}
\label{fig:boxed_graphic 6}
\end{figure}
%%%%%%%%%%%%%%%   Figure 6END   %%%%%%%%%%%%%%%%%   FFFFFFFFFFFFFFFFFFFFFFFFFFFFFFFFFFFFFF

%\clearpage

\section{Conclusion}

UHECR above GZK may be naturally born by UHE $\nu$ scattering on
relic ones. They keep, as observed, memory of distant source
direction naturally in agreement with the recent discovers of
triplets and doublets in UHECR spectra.
 The target cosmic $\nu$ may be light and dense as the needed ones in HDM model (few eV).
 Then their $W^+ W^-,ZZ$ pair productions channel and not the  Z resonant peak , would solve
the GZK puzzle. At a much lighter, but fine tuned case
$m_{\nu}\sim 0.4 eV$, $m_{\nu}\sim 1.5 eV$ assuming $E_{\nu}\sim
10^{22} eV$, one is able to solve at once the known UHECR data at
GZK edge by the dominant Z peak; in this peculiar scenario one may
foresee  (fig.2-3) a rapid   decrease (an order of magnitude in
energy fluence) above $3\cdot10^{20}eV$ in future data
   and a further recover (due to WW,ZZ channels) at higher energies.
   The characteristic UHECR fluxes will reflect the averaged
   neutrino-neutrino interactions shown in Fig.1.
   Their imprint could confirm the neutrino mass value and relic
   density. At a more extreme lighter neutrino mass, occurring for
$m_{\nu}\sim m_{\nu_SK}\sim 0.07 eV$ ,the minimal $
m_{\nu_{\tau}},m_{\nu_{\mu}}$ small mass differences might be
reflected, in a spectacular way, into UHECR modulation quite above
the GZK edges. Therefore each different neutrino mass require a
different incoming resonant  Z peak $E_{\nu}$ energy around
$3\cdot10^{20}-3\cdot10^{21} eV$ UHECR energies. These "twin"
lightest masses (Fig.4) call for either gravitational $\nu$
clustering above the expected one  \cite{Gelmini} or the presence
of relativistic diffused background. The upper bound to black
body neutrino Temperature and momentum, in a radiation dominated
Universe, is nearly $60K^o$. Such energies and comparable masses
( a few thousands of eV as the required ones in solar neutrino
puzzle) are leading to an fore-see-able scenario described by Fig
5-6. However possible gray body spectra, out of thermal
equilibrium, at higher energies may also arise from non standard
early Universe. One may be wonder if such a diffused and
homogeneous relic backgrounds are not leading by themselves to a
new $ \nu-\nu $GZK cut-off. This is not in general the case; other
obvious signatures must also be manifest \cite{Fargion2001b}.
Therefore the solution of $\nu \nu$ scattering at UHECR may probe
the real value $\nu$ density (calibrating the observed UHECR flux
intensity) revealing the known cross-sections imprint (as in Fig
1) as well as their possible lightest neutrino mass splitting
reflected in additional near future (Fig4) and (or) far future
(Fig.5-6) UHECR new knee and ankles, just near  and above GZK cut
off.  These energies are at Grand Unification edges.
 Of course the mystery of the UHECR acceleration is not yet solved,
 but their propagation from far cosmic volumes is finally allowed.
 Therefore the new generation  UHECR signature within next decade,  may offer the
best probe in testing the lightest elementary particle masses,
their relic densities and energies and the most ancient and
evasive fingerprint of $\nu$ cosmic relic background.


\begin{thebibliography}{15}

\bibitem{G} K.Greisen, 1966, Phys.Rev.Lett., 16, 748.
\bibitem{ZK} Zat'sepin, G.T., Kuz'min, V.A. 1966, JETP Lett., 4, 78
\bibitem{Proth1997} R.J.Protheroe, P.L.Biermann, 1997, Astpart.Phys 7, 181.
\bibitem{El} Elbert, J.W., Sommers, P. 1995, Apj, 441, 151
\bibitem{Sigl} P.Bhattacharjee, G.Sigl, 2000, Phys.Rept. 327,
109-247.
\bibitem{Blasi} P.Blasi, 2000, astro-ph0006316.

\bibitem{Fargion 83} Fargion,Nuovo Cimento, 77B,111, 1983 (Italy).

\bibitem{FarSal97} Fargion,A. Salis, Proc. 25th ICRC,
 Patchetstroomse,HE 4-6, p.153-156.(1997) South Africa.
\bibitem{FarSal99} D.Fargion, B.Mele, A.Salis, 1999, Astrophys. J. 517,
725.
\bibitem{Weiler} T.J.Weiler, Astropart.Phys. 11 (1999) 303-316.
\bibitem{Yoshida} S.Yoshida, G. Sigl, S. Lee, 1998, Phys.Rev.Lett. 81, 5505-5508.
\bibitem{Gelmini} G.Gelmini, 2000, hep-ph/0005263.
\bibitem{Fargion2001b} D.Fargion et all, (2001)in preparation.
\bibitem{AGASA} Y.Uchihori et al., 2000, Astropart.Phys. 13, 151-160.
%\bibitem{Kifune} T.Kifune, 1999, Astrophys.J.Lett. 518, L21.\\
%G.Amellino- Camelia et al., 1998, Nature 393, 763.
%\bibitem{Meyer} R.J.Protheroe, H.Meyer, 2000, astro-ph/0005349.
\bibitem{Konoplich} Yu. A.Golubkov, R.V. Konoplich, 1998, Phys.Atom.Nucl. 61, 602.
\bibitem{Konoplich2} D. Fargion, Yu. A. Golubkov, M. Yu. Khlopov, R. V. Konoplich, R.Mignani, 1999, JETP Lett. 69, 434-440
\bibitem{Grossi} D. Fargion, R. Konoplich, M.Grossi, M.Khlopov, 2000, Astroparticle Phys. 12, 307-314.
%\bibitem{HEGRA} D.Horns, D.Schmele, 1999, astro-ph/9909125.
%\bibitem{Milagro} J.Poirier, T.F.Lin et al., 2000,astro-ph/0004379.
\bibitem{pdg} Particle Data Group, 1996, Phys.Rev.D.
\bibitem{Enq} K. Enqvist, K.Kainulainen, J.Maalampi, 1989,Nucl.Phys. B317, 647.
\bibitem{Hill} C.T.Hill, 1983, Nucl.Phys.B224, 469.

%\bibitem{Kuzmin} O.E.Kalashev, V.A.Kuzmin, D.V.Semikoz, 2000,astro-ph/0006349.
%\bibitem{Long} Longair, 1994, {\em High energy astrophysics}, Cambridge ed.
\end{thebibliography}
\end{document}